\documentclass[12pt,a4paper]{article}
\usepackage{natbib}
\usepackage{graphicx}
\usepackage[dvips]{epsfig}
\usepackage{epic}
\usepackage{amssymb}
\usepackage{subfigure}
\usepackage{epstopdf}
\usepackage{color}
\usepackage{subeqnarray}

\newcommand{\be}{\begin{equation}}
\newcommand{\ee}{\end{equation}}
\newcommand{\bae}{\begin{eqnarray}}
\newcommand{\eae}{\end{eqnarray}}
\newcommand{\bse}{\begin{subeqnarray}}
\newcommand{\ese}{\end{subeqnarray}}
\newcommand{\pafg}[2]{\frac{\partial #1}{\partial #2}}

\begin{document}
\bigskip
\bigskip
\centerline{\Large The effect of model freshwater flux biases }
\centerline{\Large  on the multi-stable regime of the AMOC}
\bigskip \bigskip
\centerline{Henk A. Dijkstra  \footnote{Correspondence:  Email: h.a.dijkstra@uu.nl} and 
René M. van Westen}
\centerline{Institute for Marine and Atmospheric research Utrecht} 
\centerline{Department of Physics, Utrecht University} 
\centerline{Princetonplein 5, 3584 Utrecht, the Netherlands}
\bigskip
\centerline{\it Submitted to Tellus A} 
\bigskip 
\centerline{Version of \today}

\newpage
\begin{abstract}
It is known that global climate models (GCMs) have substantial biases
in the surface freshwater flux which forces the ocean  component
of these models.  
Using numerical bifurcation analyses on a global ocean model, 
we study here the effect of a specific freshwater flux bias  on the 
multiple equilibrium regime of 
the Atlantic Meridional Overturning Circulation (AMOC). We 
find that a (positive) freshwater flux bias over the Indian Ocean 
shifts the multiple equilibrium regime to larger  values of 
North Atlantic freshwater  input but hardly 
affects  the associated hysteresis width. The 
magnitude of this shift depends on the way the anomalous 
North Atlantic  freshwater  
flux is compensated.  We explain 
the changes in bifurcation diagrams   using the freshwater 
balance over the Atlantic basin. The results suggest that 
state-of-the-art GCMs may have an AMOC multiple equilibrium 
regime, but that it is located in a parameter regime that is 
considered unrealistic and hence is not explored.  
\end{abstract}

\newpage
\section{Introduction}

The AMOC has been proposed as one of the tipping elements in the climate system
\cite[]{Lenton2008,Armstrong2022}, indicating that it may undergo a relatively rapid change under 
a slowly developing forcing.  The AMOC is thought to be particularly sensitive to its freshwater 
forcing, either through the surface freshwater flux, through input of freshwater due to ice 
melt (e.g. from the Greenland Ice Sheet) or through river runoff \cite[]{Rahmstorf2005}. 
These freshwater fluxes affect the ocean density differences which control the AMOC 
strength and change the heat and salt transport carried 
by AMOC \cite[]{Marotzke2000}. 

From conceptual models,  the tipping behaviour is clearly related to the 
multi-stability properties of the  AMOC.  For example, in the Stommel 
two-box model  \cite[]{Stommel1961}, there is an interval of 
surface freshwater forcing  where two stable  steady  AMOC states exist  and tipping occurs due 
to transitions between  these states. The central feedback responsible for  the tipping 
behaviour is the salt-advection   feedback, where a  freshwater perturbation in the North  
Atlantic causes a weakening of the   AMOC which  leads to less northward salt transport  
and hence amplification of the  perturbation \cite[]{Marotzke2000}.   

Precise boundaries of the multiple-equilibrium regime of the AMOC have been obtained 
using conceptual models \cite[]{Cessi1994,Cimatoribus2012}  and fully-implicit ocean-climate models 
\cite[]{DeNiet2007,Toom2012, Mulder2021}.   One of  the important results of these studies  
is that the existence  of the multiple-equilibrium  regime can be related (in these models) to 
an observable quantity  \cite[]{Rahmstorf1996},  now often called the AMOC stability (or regime) indicator. 
This indicator has  many different notations in the literature, e.g. $M_{ov}$ \cite[]{DeVries2005} or 
$F_{ov}$ \cite[]{Hawkins2011}.  Here, we will follow \cite{Weijer2019} 
and use $F_{ovS}$  ($F_{ovN}$)  as  the  freshwater transport carried by the AMOC 
over the southern (northern) boundary at 35$^\circ$S (60$^\circ$N) of the 
Atlantic basin  \cite[]{Dijkstra2007_Tellus, Huisman2010, Liu2017}. Available 
observations \cite[]{Bryden2011}  show that the present-day AMOC is exporting freshwater 
out of the Atlantic ($F_{ovS} < 0$).  It is known that $F_{ovS}$ ignores some relevant 
processes \cite[]{Gent2018}, but if one accepts that $F_{ovS}$  is a proper indicator, 
the  AMOC is in a multiple-equilibrium  regime  based its observed values
\cite[]{Weijer2019}. 

Less precise estimates of the multiple-equilibrium regime boundaries can be obtained 
from global  climate models. In so-called  quasi-equilibrium  experiments, 
the  freshwater forcing is changed very  slowly  such that the model state stays close to 
the (slowly changing) equilibrium.   When the freshwater forcing is varied in both  
directions and covers the multiple-equilibrium  regime, regime boundaries can be inferred 
from the so-called hysteresis width, i.e.   the freshwater forcing values  where the AMOC collapses 
and recovers.  The rate  of forcing is important here and  
if this is much  faster than the equilibration time   scale of the steady  state, the approximations 
of the  regime boundaries become worse  and also rate-induced tipping may occur 
\cite[]{Lohmann2021}. Such quasi-equilibrium  experiments 
have  been performed with many  Earth System  Models of Intermediate Complexity 
(EMICs) \cite[]{Rahmstorf2005} and  the  FAMOUS model, the latter being a GCM 
with a relatively coarse horizontal resolution of  $2.5^\circ \times 3.75^\circ$ \cite[]{Hawkins2011}.  
The  Community Climate  System  Model  (CCSM3) is probably the most detailed model in 
which AMOC  hysteresis behavior has been investigated \cite[]{Hu2012}. 

Mostly due to computational constraints, often   only  
the AMOC response to particular   freshwater forcing  perturbations is considered 
in state-of-the-art GCMs. In these  so-called `hosing experiments' 
\cite[]{Stouffer2006}, quite a diversity of model behavior is found.   It is not known 
whether a multiple-equilibrium AMOC regime exists in such  models and  only sporadic indications
of such a regime have been found  \cite[]{Mecking2016, Jackson2018b, 
Jackson2018a}.  The problem is that it is difficult to assess whether the weak AMOC states 
computed are equilibrium solutions of the models. 
What these model studies certainly have shown is that an AMOC weakening would have 
severe impacts on the climate system, affecting sea level and  regional 
temperatures   in many   areas around the  world \cite[]{Vellinga2002, Jackson2015, 
Liu2020, Orihuela2022}. 

Of course, the real present-day AMOC may not have a multiple equilibrium regime and 
GCMs may model that correctly. On the other hand, the real AMOC may be in such 
a regime, and the GCMs may not capture it. In that case,  the GCMs  misrepresent  
(or miss) crucial processes, such that they do not display tipping behaviour. 
A prominent example  is the incorrect representation of  ocean eddy transport 
processes  in GCMs, which may prevent the existence of a multi-stable AMOC 
regime.  Another possibility is that  GCMs capture the relevant  processes but the 
parameters in the models are not correct, such that a multi-stable regime 
does not occur due to model biases.  An example of this  is that many GCMs are
considered to have a too stable AMOC due to  biases in the freshwater transport 
in the Atlantic Ocean \cite[]{Drijfhout2013,  Mecking2016}.   

In \cite{Westen2023}, 
it was  shown that several persistent biases in state-of-the art 
climate models,  of the Coupled Model  Intercomparison 
Projects (CMIP) version 6,   lead  to an AMOC with an Atlantic freshwater transport that 
is in disagreement with  observations (i.e., many models  have $F_{ovS} > 0$).  The 
most important model  bias is the surface freshwater  flux over the Indian Ocean, 
which affects the freshwater transport at 34$^\circ$S in the Atlantic through 
Agulhas Leakage. 
Motivated by these results, we revisit the  bifurcation analysis on the  global 
ocean-climate model \cite[]{Dijkstra2007_Tellus} and determine the effects of 
Indian Ocean surface freshwater flux biases on the multi-stable regime of the AMOC. 
In section 2, the fully-implicit model used is shortly summarised and the continuation 
methodology to compute bifurcation diagrams is  described. Then in section 3, we 
focus on the effect of  freshwater biases  on the bifurcation diagrams. Mechanisms of 
the shift in the multi-stable regimes are  analysed using the freshwater balance over the 
Atlantic. A summary and discussion follows in section 4. 

\section{Formulation}

\subsection{Model}

The  fully-implicit global ocean model used in this study  is 
described in detail in \cite{Dijkstra2005_JPO16} to which 
the reader is referred for  full details. The governing equations of the ocean 
model  are the hydrostatic,  primitive equations in spherical coordinates 
on a global domain  which includes  continental  geometry as well as 
bottom  topography.  The  ocean velocities in eastward (zonal) and northward 
(meridional)  directions are indicated by  $u$ and $v$, the vertical velocity is indicated 
by $w$, the pressure  by $p$ and the temperature and salinity by $T$ 
and $S$, respectively.  The horizontal resolution  of the model  is about 
$4^\circ$ (a $96 \times 38$  Arakawa C-grid  on a domain  
[$180^\circ$ W$,180^\circ$E] $\times$ [$85.5^\circ$ S$, 85.5^\circ$N])  and  
the grid  has 12 vertical levels. The vertical grid is  non-equidistant 
with the surface (bottom)  layer having a thickness  of 50 m  (1000 m), 
respectively. 

Vertical and horizontal mixing  of momentum and of tracers (i.e., heat and salt) are   represented by a 
Laplacian formulation  with  prescribed `eddy' viscosities $A_H$ and $A_V$ and  
eddy diffusivities $K_H$ and  $K_V$, respectively. As in \cite{Dijkstra2007_Tellus}, 
we will use the   depth  dependent  values of $K_V$ and $K_H$ \cite[]{Bryan1979, 
England1993} given by 
\bse
K_V(z) &=&   K^0_V - A_s \arctan (\lambda_V  (z - z_*)),  \\
K_H(z) &=&   K^0_H + (A_r - K^0_H) e^{\frac{z}{\lambda_H}},  
\label{e:mixK}
\ese
with  $z \in [-5000, 0]$ m.  Here, $K^0_H  = 0.5 \times  10^3$ m$^2$s$^{-1}$, 
$A_r = 1.0 \times 10^3$ m$^2$s$^{-1}$,  $K^0_V  = 8.0 \times 10^{-5}$ m$^2$s$^{-1}$, 
$A_s = 3.3  \times 10^{-5}$ m$^2$s$^{-1}$, $\lambda_V = 4.5 \times 10^{-3}$ m$^{-1}$,  
$\lambda_H = 5 \times 10^{2}$ m and $z_* = - 2.5 \times 10^{3}$ m.   A plot of the vertical structure of $K_V$ and 
$K_H$  can  be found in Figure 1 of \cite{Dijkstra2007_Tellus}. In this way, 
the vertical diffusivity $K_V$ increases  from  $0.31 \times 10^{-4}$ m$^2$s$^{-1}$ 
at the surface to   $1.3 \times 10^{-4}$ m$^2$s$^{-1}$  near the bottom 
of the flow domain.  The  horizontal diffusivity 
$K_H$   increases monotonically   from  $0.5 \times 10^{3}$  m$^2$s$^{-1}$ 
at the bottom of the ocean to  $1.0 \times 10^{3}$  m$^2$s$^{-1}$  
near the surface.  

The ocean flow is forced by the observed annual-mean wind stress as given 
in \cite{Trenberth1989}.  The upper  ocean is coupled  to a simple 
energy-balance atmospheric  model  (see Appendix in \cite{Dijkstra2005_JPO16}) 
in which only the heat transport is  modelled (no moisture 
transport).    The freshwater flux  will be prescribed in each of the  
results in section 3 and the model has no sea-ice 
component.  The surface forcing is 
represented  as a  body  forcing over the upper layer.     On the continental 
boundaries,   no-slip  conditions are prescribed and  the  heat- and salt 
fluxes are zero.   At the bottom of the ocean, both the heat   and  salt 
fluxes vanish and  slip  conditions  are assumed. 

\subsection{Methods}

The discretised  steady equations can be written as  a nonlinear 
algebraic  system of equations of the form 
\be
{\bf G}({\bf x}, \mu) = 0,
\label{e:C} 
\ee
where ${\bf x}$ is the state vector and $\mu$ is one of the parameters 
of the model.  For the  global ocean model (with a 4$^\circ$ horizontal  
resolution and   12 layers in the vertical)    the  dimension of the state 
space (and of  $\bf x$) is $96 \times 38 \times 13 \times 6 =  284,544$;
where the number $13$ comes from the $12$ ocean levels plus the 
atmospheric energy balance model. 

We use pseudo-arclength continuation \cite[]{Keller1977}, where the 
branch of steady  solutions versus $\mu$ is parametrised by an 
arclength $s$. To close the set of equations (because of the new 
variable $s$) the arclength is normalised leading to the equations
\bse
{\bf G}({\bf x}(s), \mu(s)) &=& 0, \\
\dot{\bf x}^{T}_{0} ({\bf x}(s) - {\bf x}_{0}) +  \dot{\mu}_{0}(\mu(s)  - \mu_{0}) - (s - s_{0}) &=& 0, 
\label{e:C2} 
\ese
where $({\bf x}_{0}, \mu_{0})$ is a previously computed solution and the dot indicates 
differentiation to $s$. The linear stability of each steady state is 
determined by solving a generalized eigenvalue problem using the 
Jacobi-Davidson QZ method  \cite[]{Dijkstra2005B}. 

\begin{figure}[htbp]
\begin{minipage}{\linewidth}
\centering\includegraphics[width=\linewidth]{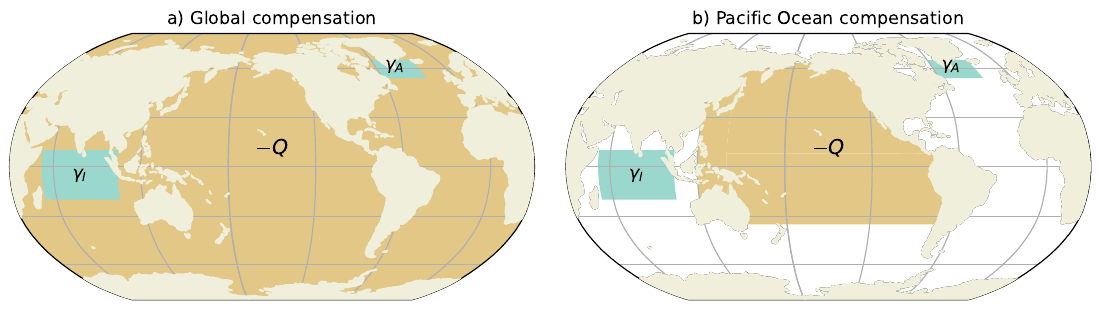} 
\end{minipage}
\caption{\emph{\small Areas  where freshwater flux anomalies are applied
with their strengths $\gamma_A$ and $\gamma_I$;  also the global 
compensation region (a) and the Pacific region (b) is shown. 
}}
\label{f:Hosing}
\end{figure}
The procedure  to compute bifurcation diagrams of the model, including biases 
in the freshwater forcing,  is the following: 
\begin{itemize} 
\item[(i)] Under restoring conditions for the surface salinity field \cite[]{Levitus1994}, 
a steady solution is determined for standard values of the parameters of the model
 \cite[]{Dijkstra2005_JPO16}.  From this steady solution the freshwater flux, below 
 referred to the Levitus flux $F^L_S$, is diagnosed. 
\item[(ii)] A  freshwater flux  over a region  near  New Foundland   (Fig.~\ref{f:Hosing}) 
with  domain [$60^\circ$ W$, 24^\circ$ W]   $\times$  [$54^\circ$ N $, 66^\circ$ N] is prescribed 
(in addition to  $F^L_S$) with  strength  $\gamma_A F^A_S$ Sv, where  $F^A_S = 1$ 
in this domain and zero outside.  Similarly,  a bias freshwater flux is prescribed 
over the  Indian Ocean domain  [$52^\circ$ E$, 104^\circ$ E] $\times$ [$20^\circ$ S, 
$10^\circ$ N] with amplitude $\gamma_I F^I_S$ Sv, where $F^I_S = 1$ in this 
domain and zero outside.  The total freshwater flux is prescribed as 
\be 
F_S = F^L_S  + \gamma_A F^A_S  + \gamma_I F^I_S - Q F^C_S, 
\label{e:fw}
\ee
where $F^C_S  = 1$ in a compensation domain (specified below) 
and the quantity  $Q$ is determined such that 
\be
 \int_{S_{oa}} F_S  ~ r^2_0 \cos \theta ~ d\theta d\phi = 0, 
\label{e:cs}
\ee
where  $S_{oa}$ is the  total ocean surface and $r_0$ the radius of the Earth.  

\item[(iii)] In the results below, we will consider two cases of compensation:
(a) global compensation, i.e. $C$ is the global ocean domain  (Fig.~\ref{f:Hosing}a)
as in \cite[]{Dijkstra2007_Tellus} and (b) $C$  is the  Pacific domain 
(Fig.~\ref{f:Hosing}b), so there is no  compensation over the Atlantic. 
In each case, for different (but fixed)  values of $\gamma_I$, a branch 
of steady solutions versus $\gamma_A$ is calculated  under the freshwater
forcing (\ref{e:fw}),  starting from  the solutions determined under (i) for 
$\gamma_A  = \gamma_I = 0$. 
\end{itemize}

\section{Results}

In the results below, we concentrate on the bifurcation diagrams and 
freshwater and salt balances. Plots of the typical AMOC patterns for 
slightly different parameter values can be found in 
\cite{Dijkstra2007_Tellus}. 

\subsection{Global Compensation}

\begin{figure}[htbp]
\begin{minipage}{\linewidth}
\centering\includegraphics[width=0.65\linewidth]{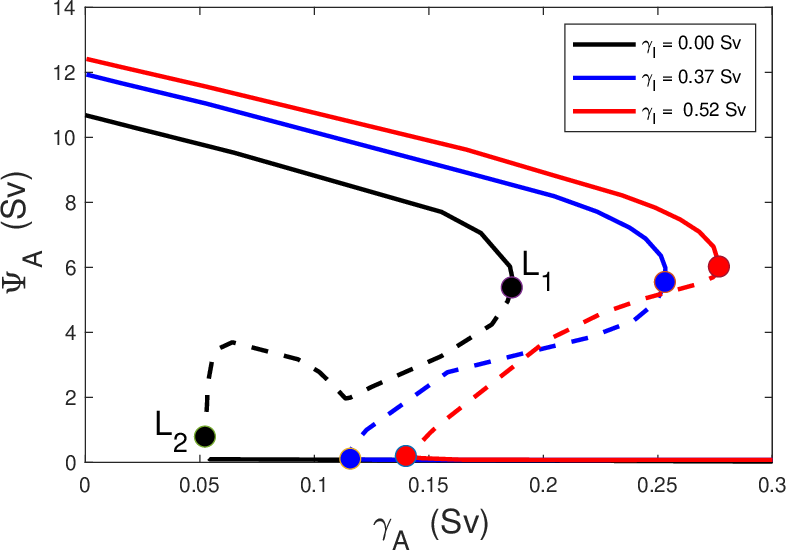} 
\end{minipage}
\caption{\emph{\small Bifurcation diagrams for the case of global compensation 
where the maximum strength of the AMOC below 1000 m   ($\Psi_A$) is plotted 
versus the strength of the  anomalous freshwater  forcing $\gamma_A$ for 
different values of $\gamma_I$. Drawn (dashed) curves indicate stable 
(unstable) branches. The dots indicate the saddle-node bifurcations. 
}}

\label{f:bif_GC}
\end{figure}

The bifurcation diagram for the global compensation case   (Fig.~\ref{f:Hosing}a) 
with $\gamma_I = 0$ (no Indian Ocean freshwater flux bias), where  the maximum  
AMOC strength below 1000 m  ($\Psi_A$) is plotted 
versus $\gamma_A$ (both in Sv), is shown as the black curve  in 
Fig.~\ref{f:bif_GC}a.  
With increasing $\gamma_A$, stable (black drawn curves) steady states exist for which 
the AMOC strength  decreases  and at $\gamma^1_A = 0.186$ Sv, a first saddle-node 
bifurcation $L_1$ occurs. With decreasing $\gamma_A$, a branch of unstable  steady states 
(black dashed curves) exists down to a second saddle-node bifurcation $L_2$ at $\gamma^2_A = 
0.054$ Sv.  

The width of the  multi-stable regime, often called  the hysteresis width $\Delta_H$,  is 
given by 
\be
\Delta_H = | \gamma^1_A - \gamma^2_A |.
\ee
For the case $\gamma_I = 0$, we find  $\Delta_H = 0.132$ Sv in this model.
In typical quasi-equilibrium  model studies \cite[]{Rahmstorf2005}, where 
$\gamma_A$ is varied with about 0.05  Sv/1000 years, 
the width is typically overestimated. With 
continuation  methods, as used here, one is able to determine the hysteresis 
width very  accurately as the values of $\gamma^{1,2}_A$  are  computed 
explicitly. 
 
With increasing values of $\gamma_I$ (adding fresh water over the Indian Ocean) 
both saddle node-bifurcations $L_1$ and $L_2$ move
to larger values of $\gamma_A$ (Fig.~\ref{f:bif_GC})  indicating that the multi-stable 
regime occurs for higher Atlantic freshwater forcing. Hence, 
with the  Indian Ocean freshwater flux bias, it is less likely that GCM having a 
`Levitus-like'  surface salinity are in a multi-stable regime (than models without 
such a bias). The width of the multi-stable regime versus $\gamma_I$ 
does not change much for these  $\gamma_I$ values; 
$\Delta_H = 0.137$ Sv and $\Delta_H = 0.138$ Sv are found for $\gamma_I = 
0.37$ Sv and $\gamma_I = 0.52$ Sv, respectively. 

To understand the shift of the branches  in the bifurcation diagram, we 
consider the Atlantic freshwater transport by the AMOC and the gyres. 
Following \cite{DeVries2005}, the quantities $F_{ov}$ (the overturning 
component) and $F_{az}$ (the azonal component) are computed as 
\bse
F_{ov}(\theta)  &=& -\frac{r_0}{S_0} \int_{S_\theta} \bar{v} (<S> - S_0) ~ dz ~ ; \\
F_{az}(\theta)  &=& -\frac{r_0}{S_0} \int_{S_\theta} \overline{v' S'} ~  dz. 
\ese
where $S_0 = 35$ psu is a reference salinity, $r_0$ is the radius of the Earth, and 
$S_\theta$ is the boundary (longitude, depth) at latitude $\theta$. Here, 
the quantities $\bar{v}$, $\bar{S}$,  $<v>$ and $<S>$ are given by 
\bse
\bar{v} &=& \int v \cos \theta ~ d\phi ~ ; ~ < v > = \frac{\bar{v}}{\int \cos \theta ~ d\phi}, \\
\bar{S} &=&  \int S \cos \theta ~ d\phi ~ ; ~ < S > = \frac{\bar{S}}{\int \cos \theta ~ d\phi}, 
\ese
and $v' = v - <v>$ and $S' = S - <S>$. The physical meaning of these quantities 
is extensively discussed in \cite{DeVries2005} and \cite{Dijkstra2007_Tellus}. 

The  existence of the saddle-node bifurcations $L_1$ and $L_2$  can be connected to the behavior of 
$F_{ovS}$. In simple box models \cite[]{Cessi1994, Rahmstorf1996}, the 
saddle-node bifurcation $L_1$ at $\gamma^1_A$ is  related to a minimum in 
$F_{ovS}$. In our model, this is more complicated as there is also a 
gyre-driven freshwater transport, and the $F_{ovS}$ minimum is only approximate. 
The saddle-node  bifurcation $L_2$ at $\gamma^2_A$ near  to a zero of 
$F_{ovS}$ along the upper branch of the AMOC, as discussed at length in  
\cite{Dijkstra2007_Tellus}.  So to explain the shift in positions in the 
saddle-node bifurcations we focus on the behaviour of $F_{ovS}$, $F_{ovN}$ 
while  also monitoring their difference $\Delta F_{ov} =  F_{ovS} - F_{ovN}$ 
and the associated behaviour of the freshwater transports by the gyres. 

\begin{figure}[htbp]
\begin{minipage}[t]{\linewidth}
\centering\includegraphics[width=0.65\linewidth]{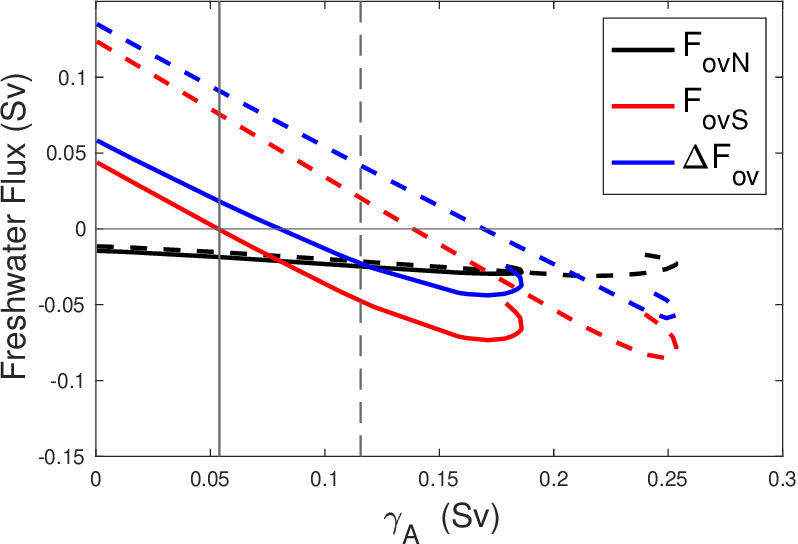} (a) 
\end{minipage}\hfill
\begin{minipage}[t]{\linewidth}
\centering\includegraphics[width=0.65\linewidth]{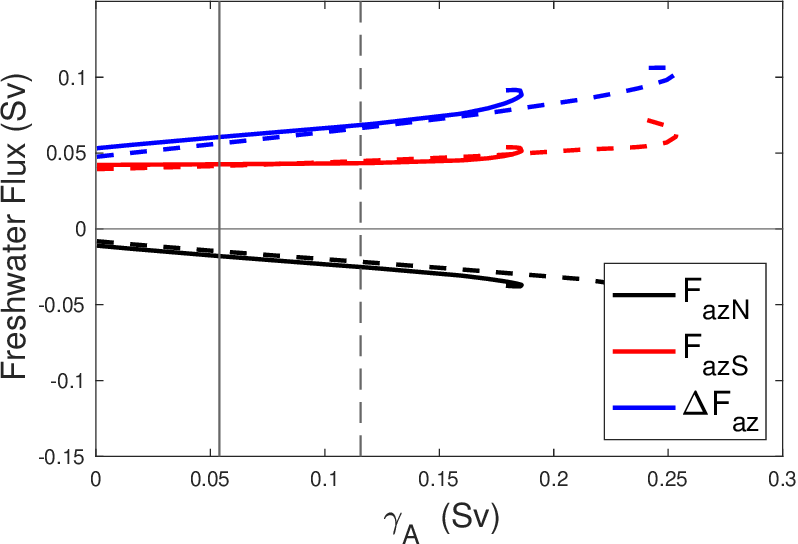} (b) 
\end{minipage}
\caption{\emph{\small Global compensation case. (a) Values of the AMOC induced 
freshwater transport at the southern boundary (35$^\circ$S) of the Atlantic ($F_{ovS}$), 
the northern 
boundary (60$^\circ$N)  of the Atlantic ($F_{ovN}$), and their difference $\Delta F_{ov}$. (b) 
Same as (a) but for the azonal transport ($F_{azS}$, $F_{azN}$ and 
$\Delta F_{az}$).  The  drawn curves are for $\gamma_I = 0$, corresponding 
to the black curve in  Fig.~\ref{f:bif_GC}.  The dashed curves are for  
$\gamma_I = 0.37$ Sv, corresponding to the blue curve in 
Fig.~\ref{f:bif_GC}. 
}}
\label{f:FovFaz_GC}
\end{figure}

The results for $F_{ovS} = F_{ov}(35^\circ$S), 
$F_{ovN} = F_{ov}(60^\circ$N) and $\Delta F_{ov}$ 
are shown in Fig.~\ref{f:FovFaz_GC}a, with the 
case $\gamma_I = 0$ ($\gamma_I = 0.37$ Sv) as  drawn (dashed) curves. 
For the chosen northern latitude (60$^\circ$N), the freshwater flux $F_{ovN}$ is negative 
and the AMOC transports freshwater  southwards  for all values of $\gamma_A$. 
For $\gamma_A = 0$, the AMOC transports freshwater  northwards at  35$^\circ$S 
as $F_{ovS} > 0$. With increasing $\gamma_A$, $F_{ovS}$ decreases and, 
for $\gamma_I = 0$,  becomes negative close to the saddle-node bifurcation $L_2$ 
at $\gamma^2_A$ (indicated by the thin drawn vertical line). For $\gamma_I = 0.37$ Sv, 
the value of $F_{ovS}$ at  $\gamma_A = 0$ is much larger (than that for 
$\gamma_I = 0$). Because $F_{ovS}$ decreases with $\gamma_A$ at the same 
rate as for $\gamma_I = 0$, the location where $F_{ovS} \approx 0$ and hence the 
bifurcation $\gamma^2_A$  occurs at a  larger value of $\gamma_A$. Also the 
location where $F_{ovS}$ obtains its minimum, and hence the position of $L_1$ 
(at $\gamma^1_A$) shifts to the right. The gyre transport changes  $F_{azS}$, 
$F_{azN}$ and $\Delta F_{az}$  with $\gamma_A$ are shown in 
Fig.~\ref{f:FovFaz_GC}b and do not change much with $\gamma_I$. 

\begin{figure}[htbp]
\begin{minipage}[t]{\linewidth}
\centering\includegraphics[width=0.7\linewidth]{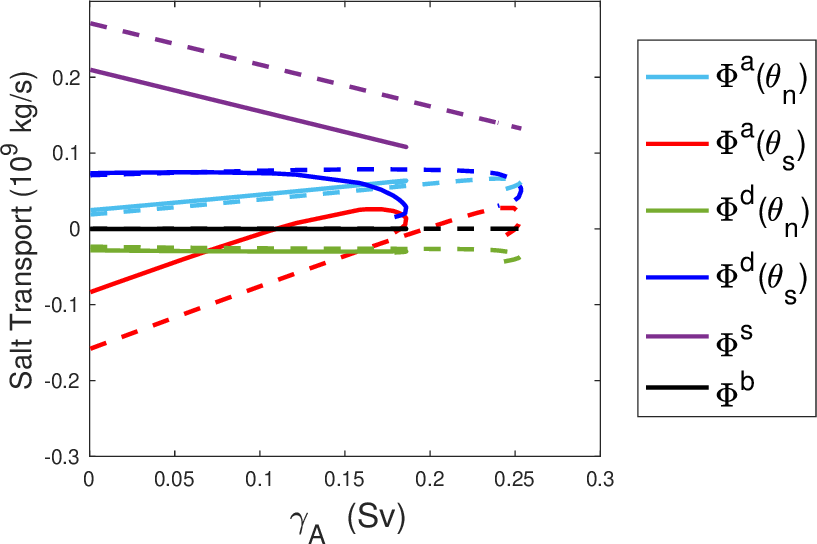} 
\end{minipage}
\caption{\emph{\small Global compensation case.  Terms in the integrated salt  balance (given in Sv) over the 
Atlantic basin over the upper branches in Fig.~\ref{f:bif_GC} (up to $\gamma^1_A$), 
with expressions  of the terms as indicated in (\ref{e:bal1a}) and (\ref{e:bal1b})  for $\theta_n = 60^\circ$N 
and $\theta_s = 35^\circ$S.  The drawn curves are  for the case $\gamma_I = 0$ and 
the dashed ones for $\gamma_I = 0.37$ Sv. 
}}
\label{f:Bal_GC}
\end{figure}

As shown in \cite{Dijkstra2007_Tellus}, the fully-implicit model allows for a closed 
salt  balance over the Atlantic from $\theta_s =35^\circ$S to $\theta_n = 60^\circ$N from 
which changes in advective, diffusive and  surface contributions can 
be determined. The terms in this  balance 
are shown in Fig.~\ref{f:Bal_GC} for both cases $\gamma_I = 0$ 
and $\gamma_I = 0.37$ Sv. Expressions for these terms were presented
in  \cite{Dijkstra2007_Tellus}, but are repeated here for convenience, i.e. 
\bse
\Phi^s  &=& \int_{S_{oa}} S_0 F_S  ~ r^2_0 \cos \theta ~  d \phi d\theta,    \\
\Phi^a(\theta)  &=& -  \int_{S_\theta} v S ~   r_0 \cos \theta ~ d \phi  dz,  \\
\Phi^d(\theta) &=&  \int_{S_\theta}  K_H  \pafg{S}{\theta}  ~  \cos \theta ~ d \phi  dz,  
\label{e:bal1a} 
\ese
where $\Phi^a$ and $\Phi^d$  are the  advective and diffusive fluxes through the 
boundary $S_\theta$, respectively. The overall balance is given by 
\be
\Phi^b = \Phi^a(\theta_n) - \Phi^d(\theta_n) - \Phi^a(\theta_s) + \Phi^d(\theta_s) -\Phi^s 
\label{e:bal1b} 
\ee
Indeed, the term $\Phi^b$ is much smaller than the individual terms
(black curves in Fig.~\ref{f:Bal_GC})  giving a nearly closed salt  balance 
over the Atlantic basin for all values of the parameters. 

First consider the case $\gamma_I = 0$ (drawn curves in Fig.~\ref{f:Bal_GC})
and the upper branch in the bifurcation diagram up to $L_1$. 
For $\gamma_A = 0$, 
the surface (virtual) salt   flux  is approximately balanced by the fluxes 
at the southern boundary. The  surface evaporation is larger than the 
precipitation ($\Phi^s > 0$) and this salt is transported out of the Atlantic 
basin at the southern boundary  ($\Phi^a(\theta_s) < 0$).  The fact  that the 
diffusive flux is relatively large here  (compared to typical GCMs)  is 
that the model has a  coarse resolution so needs a relatively high horizontal  
diffusivity to prevent wiggles to occur near the boundaries. The salt 
fluxes at the northern boundary are less important  and the respective 
components are about a factor 2 to 4 smaller than at the southern boundary. 

With increasing $\gamma_A$, the surface salt flux decreases as freshwater 
is put into the  North Atlantic. The diffusive salt transports do not 
respond but the southward advective salt transport at $\theta_s$ 
weakens and eventually changes sign near $\gamma_A = 0.1$ Sv. As 
the diffusive flux is directed to 
transport salt into the basin at the southern boundary, the value of 
$\gamma_A$ where the sign change in salt transport occurs is 
around $0.06$ Sv (Fig.~\ref{f:Bal_GC}). Note
that the gyre and AMOC components cannot be distinguished in the 
advective fluxes. 

When $\gamma_I = 0.37$ Sv (dashed curves in Fig.~\ref{f:Bal_GC}), 
the surface salt flux increases for  $\gamma_A = 0$ compared to the 
case $\gamma_I = 0$. This is due to the global compensation as a 
negative salt flux in the Indian Ocean is compensated by a 
positive one over part of the Atlantic. Hence, the curve for $\Phi^s$ 
shifts upwards and so the compensating advective flux at 
the southern boundary shifts downwards. A second effect
is that the changed surface freshwater flux pattern leads to a modified 
salinity  distribution in the Atlantic. This   increases 
the AMOC (Fig.~\ref{f:bif_GC}) strength and hence also its
salt transport out of  the basin at the southern boundary. 

Because the diffusive fluxes
are not much affected by the Indian Ocean freshwater input, the 
fluxes $\Phi^s$ and  $\Phi^a(\theta_s)$ change with $\gamma_A$ 
in the same way as for the case $\gamma_I = 0$. The  
starting value of $\Phi^s$ at $\gamma_A = 0$ is now larger and 
it takes a larger value of $\gamma_A$ to change the sign of 
the freshwater flux at the southern boundary and to reach a
minimum in this quantity. Hence the  saddle-node bifurcations  
$L_1$ and $L_2$ shifts to larger values of $\gamma_A$. 

\subsection{Pacific Compensation} 

\begin{figure}[htbp]
\begin{minipage}{\linewidth}
\centering\includegraphics[width=0.65\linewidth]{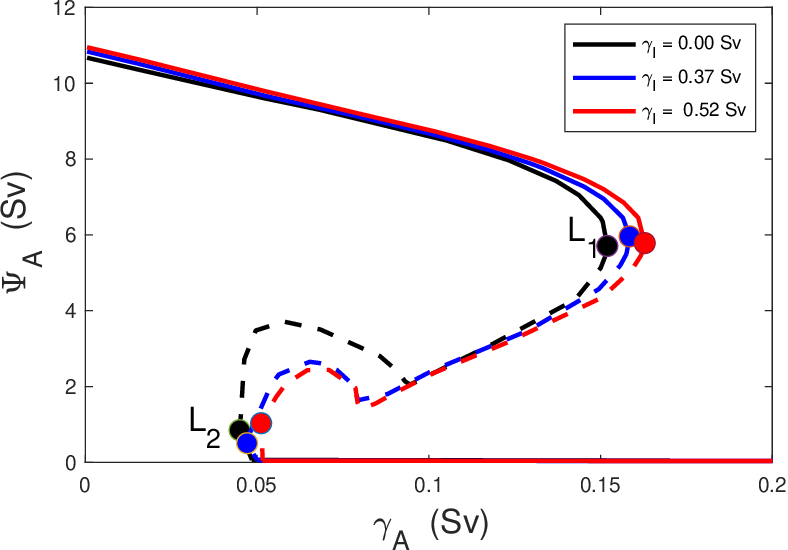} 
\end{minipage}
\caption{\emph{\small Bifurcation diagram for the case of Pacific compensation 
where the maximum strength of the AMOC below 1000 m  ($\Psi_A$) is plotted 
versus the strength of the 
anomalous Atlantic freshwater  forcing $\gamma_A$ for different values of 
$\gamma_I$.  The dots indicate the saddle-node bifurcations. 
}}
\label{f:bif_PC}
\end{figure}

Since  the global compensation  has a substantial influence 
on the position of the saddle-node bifurcations  (Fig.~\ref{f:bif_GC}), 
we now consider the case where compensation is only over the Pacific 
domain as indicated in Fig.~\ref{f:Hosing}b. The bifurcation diagrams in 
this case are, for different values of $\gamma_I$, shown in 
Fig.~\ref{f:bif_PC}. The shift of the saddle-nodes to larger values 
of $\gamma_A$  is much smaller than for the global compensation case 
(Fig.~\ref{f:bif_GC}). The hysteresis width itself for  $\gamma_I = 0$  
with $\Delta _H = 0.103$ Sv for the Pacific compensation case is a 
bit smaller than for the global compensation case ($\Delta _H = 0.138$ Sv). 
This width is only slightly larger for the case $\gamma_I = 0.52$ Sv, i.e. 
$\Delta _H = 0.111$ Sv. 

\begin{figure}[htbp]
\begin{minipage}[t]{\linewidth}
\centering\includegraphics[width=0.65\linewidth]{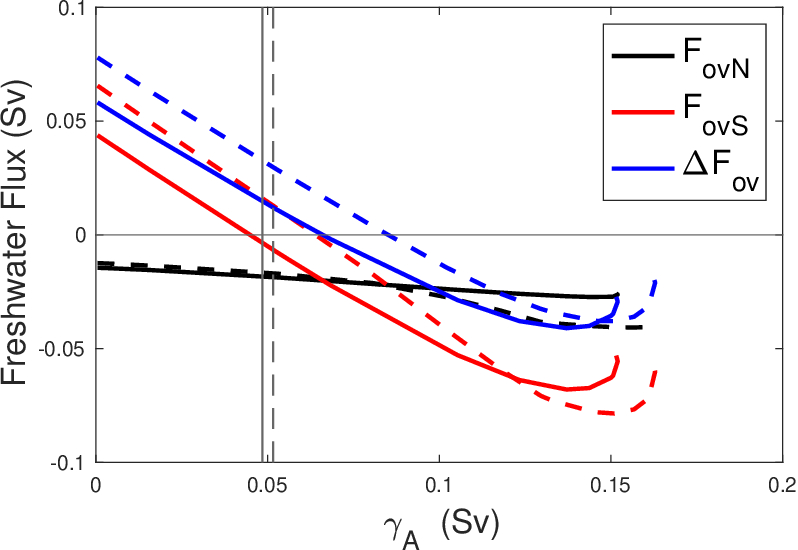} (a) 
\end{minipage}\hfill
\begin{minipage}[t]{\linewidth}
\centering\includegraphics[width=0.65\linewidth]{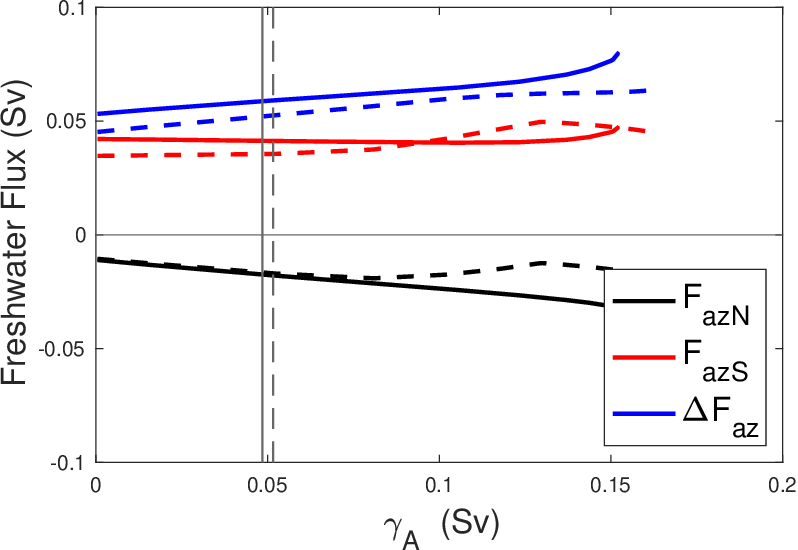} (b) 
\end{minipage}
\caption{\emph{\small Pacific compensation case. (a) Values of the AMOC induced freshwater transport 
at the southern boundary (35$^\circ$S) of the Atlantic ($F_{ovS}$), the northern 
boundary (60$^\circ$N)  of the Atlantic ($F_{ovN}$), and their difference ($\Delta F_{ov}$). (b) 
Same as (a) but for the azonal transport ($F_{azS}$, $F_{azN}$ and 
$\Delta F_{az}$).  The 
drawn curves are for $\gamma_I = 0$, corresponding to the black curve in 
Fig.~\ref{f:bif_PC}.  The dashed curves are for  $\gamma_I = 0.52$ Sv, 
corresponding to the red curves in Fig.~\ref{f:bif_PC}. 
}}
\label{f:FovFaz_PC}
\end{figure}

For the analysis of the freshwater and salt balances, we choose the 
larger value ($\gamma_I = 0.52$ Sv) instead of $\gamma_I = 0.37$ Sv 
(used in the global compensation case), as the differences are more 
clearly visible. 
The freshwater transports by the AMOC and by the gyres 
(Fig.~\ref{f:FovFaz_PC})  show that for $\gamma_A = 0$, 
$F_{ov}$ is larger for $\gamma_I = 0.52$ Sv (Fig.~\ref{f:FovFaz_PC}a), 
compared to the case  $\gamma_I = 0.0$ Sv. 
Hence also changes in the freshwater balance are induced in the Atlantic, 
but both the direct effect of compensation of the surface freshwater flux 
and the secondary effect of an AMOC increase are much smaller. 
Of course, in such a diffusive and viscous model, the Agulhas 
retroflection is in a diffusive retroflection regime \cite[]{Dijkstra2001_JPO4} 
but there is additional  freshwater  transport from the Indian to the 
Atlantic when $\gamma_I > 0$.  As the wind-driven 
freshwater transport does not change  much with $\gamma_A$ 
(Fig.~\ref{f:FovFaz_PC}b), the AMOC transports  more freshwater into 
the basin  and hence $F_{ovS}$  becomes more positive. 

\begin{figure}[htbp]
\begin{minipage}[t]{\linewidth}
\centering\includegraphics[width=0.7\linewidth]{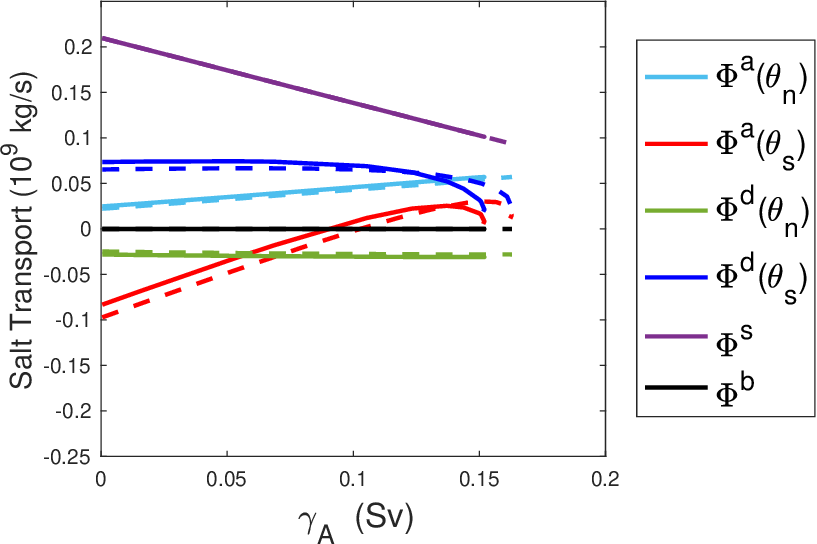} 
\end{minipage}
\caption{\emph{\small Pacific compensation case. Terms in the integrated salt balance over the 
Atlantic basin over the upper branch in Fig.~\ref{f:bif_PC}, with expressions 
of the terms as indicated in the main text.  The drawn curves are for 
the case $\gamma_I = 0$ and the dashed ones for 
$\gamma_I = 0.52$ Sv. 
}}
\label{f:Bal_PC}
\end{figure}

In the Atlantic salt balance (Fig.~\ref{f:Bal_PC})   the surface 
salt flux is indeed the same for both values of $\gamma_I$, because 
there is no compensation anymore over the Atlantic. The advective 
transport of salt becomes  slightly more negative over the southern 
boundary for $\gamma_I = 0.52$ Sv indicating that  indeed a small 
amount of freshwater is  transported into the Atlantic basin. Also the 
diffusive transport of  salt (into the basin) decreases  and this 
approximately  balances the advective contribution. As the advective 
contribution is much smaller than the compensation contribution 
in the global compensation case, the shift of the saddle-node bifurcations
to larger values of $\gamma_A$ is much smaller. 

\section{Summary and Discussion}

Following earlier studies on CMIP3 and CMIP5 models  \cite[]{Drijfhout2013,  Mecking2016}, 
also many CMIP6 models  have large biases in surface freshwater fluxes which lead 
to an AMOC  with an Atlantic freshwater  transport that is in disagreement  
with observations  \cite[]{Westen2023}. The most important model  bias 
is in the Atlantic Surface  Water properties, which arises from deficiencies in the surface 
freshwater  flux over the Indian Ocean. A second bias is in the properties in the North 
Atlantic Deep Water and  arises through deficiencies in the  freshwater flux over the 
Atlantic Subpolar Gyre region \cite[]{Westen2023}. 

In this paper, we have addressed to effects of the freshwater flux bias in the Indian 
Ocean  on the multiple equilibrium regime  of the AMOC using the fully implicit global 
ocean-atmosphere model  of \cite{Dijkstra2005_JPO16}, for which explicit bifurcation 
diagrams can be computed. This is a fantastic capability as both stable and unstable 
steady states can be computed, and the width of the multiple equilibrium regime can
be determined accurately. However, this can only be done with quite a simplified 
global model, with  relatively  low resolution and  hence being more viscous and diffusive than 
current, even low-resolution,  ocean models. The atmosphere  model is only an energy balance model 
with a prescribed freshwater forcing. In terms of quantitative results on changes 
of the bifurcation diagrams, the model is probably not that useful. 

Qualitatively, however, the model provides very useful information, as it indicates 
that the multiple equilibrium regime does not disappear due to the Indian 
Ocean freshwater flux biases, but that it shifts  to higher values of the 
North Atlantic anomalous freshwater  flux. This shift is  dependent on the
way the latter flux is compensated. There are two mechanisms which 
are responsible for these shifts: surface salinity patterns in the Atlantic 
depend on the compensation of the Indian Ocean bias and, in  both cases 
considered here, lead to a slight  increase in  the AMOC. In 
the Levitus background  state (with $F_{ovS} > 0$), this leads to a larger transport of salt 
out of the Atlantic basin. Hence, a larger anomalous North Atlantic surface freshwater flux  is needed  
to activate  the salt advection feedback,  i.e., to a situation   where the  AMOC exports 
fresh water. When there is compensation of the Indian Ocean freshwater 
flux in the Atlantic, the Atlantic becomes saltier and hence the AMOC 
transports more salt out of the basin. This  leads to an additional, and 
here  larger,  shift of the bifurcation diagram compared to when there 
is no compensation in the Atlantic. 

The mechanisms identified are  
useful for interpreting results from GCMs and for designing new 
simulations with these models.  First it shows that biases in freshwater flux 
lead to shifts in the bifurcation diagram which would imply that 
such multiple equilibrium regimes (and hence AMOC collapses) 
could exist in these models but are located in a parameter 
regime where one would not normally  perform simulations. Second, 
it provides a hint why  efforts  to find AMOC collapses in these 
models may not have been successful \cite[]{Mecking2016, Jackson2018a, 
Jackson2018b}. Even an enormous  freshwater input in a parameter regime 
which is outside the  multiple equilibrium regime would only lead to a 
weakened (but no collapsed) AMOC. 

This also suggests several ways forward to find an AMOC collapse 
in state-of-the-art models. One either performs a long quasi-equilibrium
simulation up to very large freshwater flux input such as in the FAMOUS 
model \cite[]{Hawkins2011} to find the collapse 
(this may take a few thousand years of simulation, so is  expensive). In doing 
this, it is better to compensate  outside of the Atlantic 
when using surface fluxes, as the latter will introduce an additional 
shift and so one has to integrate longer to find an AMOC collapse. Such 
compensation procedures  (including compensation over the volume) 
are now also more common  in  GCMs \cite[]{Jackson2022}.  The 
alternative is to address and improve  the biases in the 
atmospheric components of the models, but this is not an easy issue. 
The origin of these biases may even   be a coupled problem, as the bias 
strength is positively correlated with the AMOC strength 
\cite[]{Westen2023}. 

We hope  that this study will motivate  the design of new AMOC hosing 
simulations  to find AMOC collapses in state-of-the-art GCMs. Detection 
of such a collapse  would have a large impact on climate change 
research and  probability estimates  of AMOC tipping under global 
warming  would likely need to  be revised. 

\vfill 
\noindent{\Large Acknowledgements}

\noindent  H.A.D. and R.M.v.W. are funded by the European Research 
Council through the ERC-AdG project  TAOC (project 101055096). The
authors thank Dr. Fred Wubs (University of Groningen, NL) for helping
with revising the old Fortran code of the model 
\cite[]{Dijkstra2007_Tellus} used in this paper. 

\newpage
\bibliography{tellus2v_r1, erc_2021_total,References} 

\end{document}